\documentclass[fdp-class,a4paper,fleqn%
]{w-art-class}
\usepackage{times,cite,w-thm,amssymb,amsfonts,amsmath}
\usepackage{bbm,array,graphicx,wrapfig,float,mathtools,slashbox,multirow}
\newcommand{\be}{\begin{equation}}
\newcommand{\ee}{\end{equation}}
\newcommand{\beq}{\begin{equation}}
\newcommand{\beql}[1]{\begin{equation}\label{#1}}
\newcommand{\eeq}{\end{equation}} 
\newcommand{\ba}{\begin{array}}
\newcommand{\ea}{\end{array}}
\newcommand{\bea}{\begin{eqnarray}}
\newcommand{\beal}[1]{\begin{eqnarray}\label{#1}}
\newcommand{\eea}{\end{eqnarray}}
\newcommand{\ben}{\begin{enumerate}}
\newcommand{\een}{\end{enumerate}}
\newcommand{\bean}{\begin{eqnarray*}}
\newcommand{\eean}{\end{eqnarray*}}

\newcommand{\btab}[1]{\begin{tabular}{#1}}
\newcommand{\etab}{\end{tabular}}

\newcommand{\Diag}{\mbox{Diag}}
\newcommand{\bmi}{\begin{minipage}}
\newcommand{\emi}{\end{minipage}}

\newcommand{\BC}{\mathbb{C}}

\newcommand{\BZ}{\mathbb{Z}}
\newcommand{\comment}[1]{}

\usepackage[]{graphicx}
\begin{document}
%%    The information for the title page will be placed between
%%    \begin{document} and \maketitle. The order of most entries
%%    is determined by the class file and can not be changed by
%%    rearranging them. The maketitle command follows after the
%%    abstract.
%%
%%    Most of the following commands will be completed by the publisher.
%%
%%    The copyrightyear is defined in the .clo file as the first argument
%%    of the copyrightinfo command. If the copyrightyear differs from that
%%    value it might be adjusted by the following definition:
%%
%% \renewcommand{\copyrightyear}{2007}% uncomment to change the copyrightyear.
%%
%\DOIsuffix{theDOIsuffix}
%%
%% issueinfo for the header line
%\Volume{55}
%\Month{01}
%\Year{2007}
%%
%%    First and last pagenumber of the article. If the option
%%    'autolastpage' is set (default) the second argument may be left empty.
\pagespan{1}{}
%%
%%    Dates will be filled in by the publisher. The 'reviseddate' and
%%    'dateposted' (Published online) entry may be left empty.
%\Receiveddate{XXXX}
%\Reviseddate{XXXX}
%\Accepteddate{XXXX}
%\Dateposted{XXXX}
%%

%% \pretitle{Editor's Choice}

%% We have a short and a long form for the title. The short form
%% (optional argument) goes into the running head.

\title[An Introduction to Counting Orbifolds]{An Introduction to Counting Orbifolds}
%% Please do not enter footnotes or \inst{}-notes into the optional
%% argument of the author command. The optional argument will go into
%% the header.  If there is only one address the marker \inst{x} may be
%% omitted.

%% Information for the first author.
\author[John Davey]{John Davey\inst{1}%
  \footnote{\quad E-mail:~\textsf{jpdavey@imperial.ac.uk},
            }}
\address[\inst{1}]{Theoretical Physics,
Blackett Laboratory,
Imperial College London,
London, SW7 2AZ}
%%
%%    Information for the second author
\author[Amihay Hanany]{Amihay Hanany\inst{1}
 \footnote{\quad E-mail:~\textsf{a.hanany@imperial.ac.uk},
            }}
%%
%%    Information for the third author
\author[Rak-Kyeong Seong]{Rak-Kyeong Seong\inst{1}
 \footnote{\quad E-mail:~\textsf{rak-kyeong.seong@imperial.ac.uk},
            }}
%%
%%    \dedicatory{This is a dedicatory.}
\begin{abstract}
We review three methods of counting abelian orbifolds of the form $\BC^3/\Gamma$ which are toric Calabi-Yau (CY). The methods include the use of 3-tuples to define the action of $\Gamma$ on $\BC^3$, the counting of triangular toric diagrams and the construction of hexagonal brane tilings. A formula for the partition function that counts these orbifolds is given. Extensions to higher dimensional orbifolds are briefly discussed.
\end{abstract}
%% maketitle must follow the abstract.
\maketitle                   % Produces the title.

%% If there is not enough space inside the running head
%% for all authors including the title you may provide
%% the leftmark in one of the following three forms:

%% \renewcommand{\leftmark}
%% {First Author: A Short Title}

%% \renewcommand{\leftmark}
%% {First Author and Second Author: A Short Title}

%% \renewcommand{\leftmark}
%% {First Author et al.: A Short Title} 

%% \tableofcontents  % Produces the table of contents.
\section{Introduction}

Orbifolds have played a pivotal role in both mathematics and string theory. Born in mathematics from discussions of manifolds and quotient spaces \cite{Satake,Thurston}, orbifolds were embraced as an avant-garde subject in string theory. The key advance was the idea of compactifying string theory on orbifolds \cite{preDixon,Dixon}; this has been promptly recognized as offering new possibilities to the community. A myriad of work and interest followed, ranging from applications in conformal field theories \cite{Dixon:1986qv} and heterotic string theory \cite{Ibanez:1987pj}, to cosmic strings \cite{Greene:1989ya}.

More recently, orbifolds gained prominence through the subject of brane resolutions of Calabi-Yau moduli spaces. D3-branes which probe non-compact abelian orbifolds of $\mathbb{C}^{3}$ \cite{DouglasMoore96,DouglasMoore97,DouglasGreeneMorrison97,Muto:1997pq,Beasley:1999uz} have a world volume theory which is a $(3+1)$-dimensional quiver gauge theory \cite{Klebanov:1998hh,Acharya:1998db}. In M-theory, works by Bagger-Lambert \cite{BaggerLambert07,BaggerLambert08a,BaggerLambert08b}, Gustavsson \cite{Gustavsson07,Gustavsson08} and Aharony-Bergman-Jafferis-Maldacena (ABJM) \cite{ABJM08}, led to the investigation of M$2$-branes which probe orbifolds of $\mathbb{C}^{4}$. The world volume theory of M$2$-branes on orbifolds is a $\mathcal{N}=2$ $(2+1)$-dimensional quiver Chern-Simons theory \cite{Martelli:2008si,Hanany:2008cd,Hanany:2008gx}.

Looking back at the past study of orbifolds, one notices that there have been relatively few systematic studies on enumerating orbifolds. For instance, taking branes on orbifold singularities, it is widely known that there are two abelian orbifolds of the form $\BC^3/\Gamma$ at order $|\Gamma|=3$, which are $\BC^3 / \BZ_3$ -- sometimes known as the cone over $\mathrm{dP}_0$ -- and $\BC^2 / \BZ_2 \times \BC$. An unanswered question has been how many distinct abelian orbifolds of $\mathbb{C}^3$ there are for an arbitrary order of $\Gamma$.

We review the systematic study of abelian orbifolds of the form $\mathbb{C}^d / \Gamma $ with $\Gamma$ being a finite abelian subgroup of $SU(d)$. We count these orbifolds according to the order of the group $\Gamma$. These orbifolds are toric Calabi-Yau (CY) singularities. Particular attention shall be drawn to the case $d=3$.

The three methods of counting orbifolds which are discussed are:
\begin{itemize}
\item Using 3-tuples that specify actions of the generators of $\Gamma$ on $\BC^3$. There are some technical details which make this approach difficult. Full details of this method are given in Section \ref{s:OrbEnum}
\item
Exploiting the toric description of abelian orbifolds. Abelian orbifolds of $\mathbb{C}^{3}$ correspond to triangles on a $\BZ^2$ lattice. The counting of orbifolds using this method is covered in Section \ref{s:ToricEnum}.
\item
Counting all possible Brane Tilings that can be constructed using only hexagons. A Brane Tiling (or Dimer Model) is a graphical representation of the world volume theory of a D3-brane that probes a toric singularity \cite{HananyKen05,Hanany05,Yamazaki:2008bt,Davey:2009bp,Hanany:2005ss,Franco:2005sm,Kennaway:2007tq}. Brane tilings formed from only hexagons have a moduli space which is an abelian orbifold of $\BC^3$. The details of this method can be found in Section \ref{s:TilingEnum}.
\end{itemize}

All three of the methods above are found to give an identical counting of orbifolds of the form $\BC^3 / \Gamma$. The counting is given explicitly in Section \ref{s:Counting}. A formula for the partition function that counts these orbifolds is also given \cite{HananyOrlando10}. A discussion on generalizing the above methods to count higher dimensional orbifolds of the form $\BC^d/\Gamma$ for $d>3$ is given. Full details of the methods described in this review can be found in \cite{Orbs}.

%%%%%%%%%%%%%%%%%%%%%%%%%%%%%%%%%%%%%%%%%%%%%%%%%%%%%%%%%%%%%%%%%%%%%%%%%%%%%%%%%%%%%%%%%%%%%%%%%%%%%%%%%%%%%%%%%%%%%%%%%%%%
\section{Counting Orbifolds Using 3-tuples}
\label{s:OrbEnum}
A method of counting orbifolds of $\mathbb{C}^{3}$ using a collection of 3-tuples is described in this section. Let us consider the quotient formed when $\Gamma$, a finite abelian subgroup of $SU(3)$, acts on the space $\BC^3$. The resulting space is a toric non-compact Calabi-Yau (CY) singularity. 

As the group $\Gamma$ is abelian, the most general form can be written as the product $\Gamma = \BZ_{n_1} \times \BZ_{n_2}$ with $|\Gamma| = n_1 n_2$. Let $g$ be a generator of one of the $\BZ_{n_i}$. Then the corresponding representation is given by
\beq
g ~~=~~
\left( 
\begin{array}{ccc}
e^{\frac{i2\pi a_1}{n_i}} & 0 & 0\\
0&e^{\frac{i2\pi a_2}{n_i}} & 0 \\
0&0&e^{\frac{i2\pi a_3}{n_i}} \\
\end{array}
\right)
~~=~~\Diag \left(
e^{\frac{i2\pi a_1}{n_i}} ,
e^{\frac{i2\pi a_2}{n_i}} , 
e^{\frac{i2\pi a_3}{n_i}} 
\right)
\eeq
The action of the group $\BZ_{n_i}$ is therefore encoded by three integer parameters $a_i$ which satisfy $(a_1+a_2+a_3)=0 ~~(\bmod \; {n_i})$. We can keep track of this action in a 3-tuple $(a_1,a_2,-a_1-a_2)$. A list of these 3-tuples, each defining an action for a $\BZ_{n_i}$, can be used to define an orbifold action for a composite group $\Gamma$.

One way in which one can count orbifolds is to simply consider all possible collections of 3-tuples that can form an action. One must then take into account that the same geometry could be defined by two different collections of 3-tuples.

\subsection{Over-counting Issues}
\label{s:OC}
There are different ways in which a set of 3-tuples that define an orbifold action can give rise to the same geometry. Let the following summarize the ambiguity:
\begin{itemize}
 \item There is a freedom of choosing the parameterization of $\BC^3$ by the coordinates $z_i$. One should consider two quotients equivalent if they are related to each other by a permutation of these coordinates.
\item The generators of each $\BZ_{n_i}$ are not necessarily unique. For instance, if one considers a generator $g \in \BZ_5$ then $g^2$, $g^3$ and $g^4$ are all generators of the group $\BZ_5$. Therefore if one has a 3-tuple $(a_1 , a_2 , a_3)$ that defines the action of some group $\BZ_n$ on $\BC^3$ then, for $\lambda$ co-prime to $n$, the 3-tuple $\lambda (a_1 , a_2 , a_3)$ defines an equivalent orbifold action. The convention used here is to only consider 3-tuples $(a_1,a_2,a_3)$ that satisfy $\mathrm{gcd}(a_1,a_2,a_3) =1$.
\item If $p$ and $q$ are co-prime, $\BZ_p \times \BZ_q = \BZ_{pq}$. Therefore orbifolds of composite order can be equivalent to orbifolds formed by a single $\mathbb{Z}_n$ acting on $\mathbb{C}^3$.
\end{itemize}
\subsection{An Example - $\BC^3 / \BZ_3$}
To explicitly illustrate some of the issues that are discussed above, let us consider the example of abelian orbifolds of the form $\BC^3 / \Gamma$ for $|\Gamma|=3$. The only abelian subgroup of $SU(3)$ of order $3$ is $\BZ_3$. By enumerating all 3-tuples that correspond to orbifolds actions of $\BZ_3$, one finds that there are 7 such 3-tuples. These are given in Table \ref{tab:3vecs}. After consideration of the over-counting issues given in Section \ref{s:OC}, it can be deduced that there are $2$ distinct abelian orbifolds of $\BC^3$ at order $3$. One orbifold has the orbifold action $(0,1,2)$ and is known in the literature as $\BC^2/\BZ_3 \times \BC$. The other orbifold has the action $(1,1,1)$ and is often referred to as $\BC^3 / \BZ_3$ or as the cone over the del Pezzo 0 $(\mathrm{dP}_0)$ surface.

\begin{table}
\bmi{2.25in}
\begin{tabular}{|c|c|}
\hline
{\bf Orbifold Name} & {\bf Orbifold Action}\\
\hline
\multirow{6}{*}{$\BC^2/\BZ_3 \times \BC$} & $(0,1,2)$\\
 & $(0,2,1)$\\
 & $(1,0,2)$\\
 & $(2,0,1)$\\
 & $(1,2,0)$\\
 & $(2,1,0)$\\
\hline
\end{tabular}
\emi
\bmi{2in}
\begin{tabular}{|c|c|}
\hline
{\bf Orbifold Name} & {\bf Orbifold Action} \\
\hline
$\BC^3 / \BZ_3$ & $(1,1,1)$\\
\hline
\end{tabular}
\emi
\caption{The two distinct orbifolds of the form $\mathbb{C}^3/\Gamma$ at order $|\Gamma|=3$. \label{tab:3vecs}}
\end{table}

\subsection{Consideration of $\BC^3 / (\BZ_n \times \BZ_m)$}
When considering orbifolds corresponding to groups of composite order, two 3-tuples must be used to keep track of the orbifold action of the abelian product group. A detailed discussion for this case is given in \cite{Orbs}.
%%%%%%%%%%%%%%%%%%%%%%%%%%%%%%%%%%%%%%%%%%%%%%%%%%%%%%%%%%%%%%%%%%%%%%%%%%%%%%%%%%%%%%%%%%%%%%%%%%%%%%%%%%%%%%%%%%%%%%%%%%%%
\section{Counting Orbifolds using the Toric Description}
\label{s:ToricEnum}
A second way in which it is possible to count abelian orbifolds of $\BC^3$ is to use their toric description \cite{Fulton}. A toric Calabi-Yau 3-fold can be represented by a convex polygon in a $\mathbb{Z}^{2}$ lattice. Two such polygons correspond to the same manifold if and only if they are related to each other by a $GL(2 , \BZ)$ transformation. Abelian orbifolds of $\mathbb{C}^{3}$ are toric and have lattice triangles as their toric diagrams. Therefore it is possible to count distinct abelian orbifolds of $\mathbb{C}^3$ by considering all triangles in a $\BZ^2$ lattice that are not related to each other by a $GL(2 , \BZ)$ transformation.

The area of a toric triangle in $\BZ^2$ equals the order of the group, $| \Gamma |$, in $\BC^3 /  \Gamma$. Therefore, to count orbifolds according to $|\Gamma|$, all toric triangles of area $|\Gamma|$ must be generated first. This can be done by multiplying each of the vectors that represent the vertices of a unit triangle by $2 \times 2$ integer valued matrices of determinant $|\Gamma|$.

As an example, it is possible to generate triangles of area $2$ by using integer valued $2 \times 2$ matrices of determinant $2$. One could multiply each of the vectors $\{ {0 \choose 0} , {1 \choose 0 }, {0 \choose 1} \}$ by the matrix  ${1 \; 0 \choose 0 \; 2} $ to get the vectors $\{ {0 \choose 0} , {1 \choose 0 }, {0 \choose 2} \}$ which corresponds to a triangle of area 2 in a $\BZ^2$ lattice,
\beq
\includegraphics[height=0.7cm, trim=0cm 3cm 0cm 0cm ]{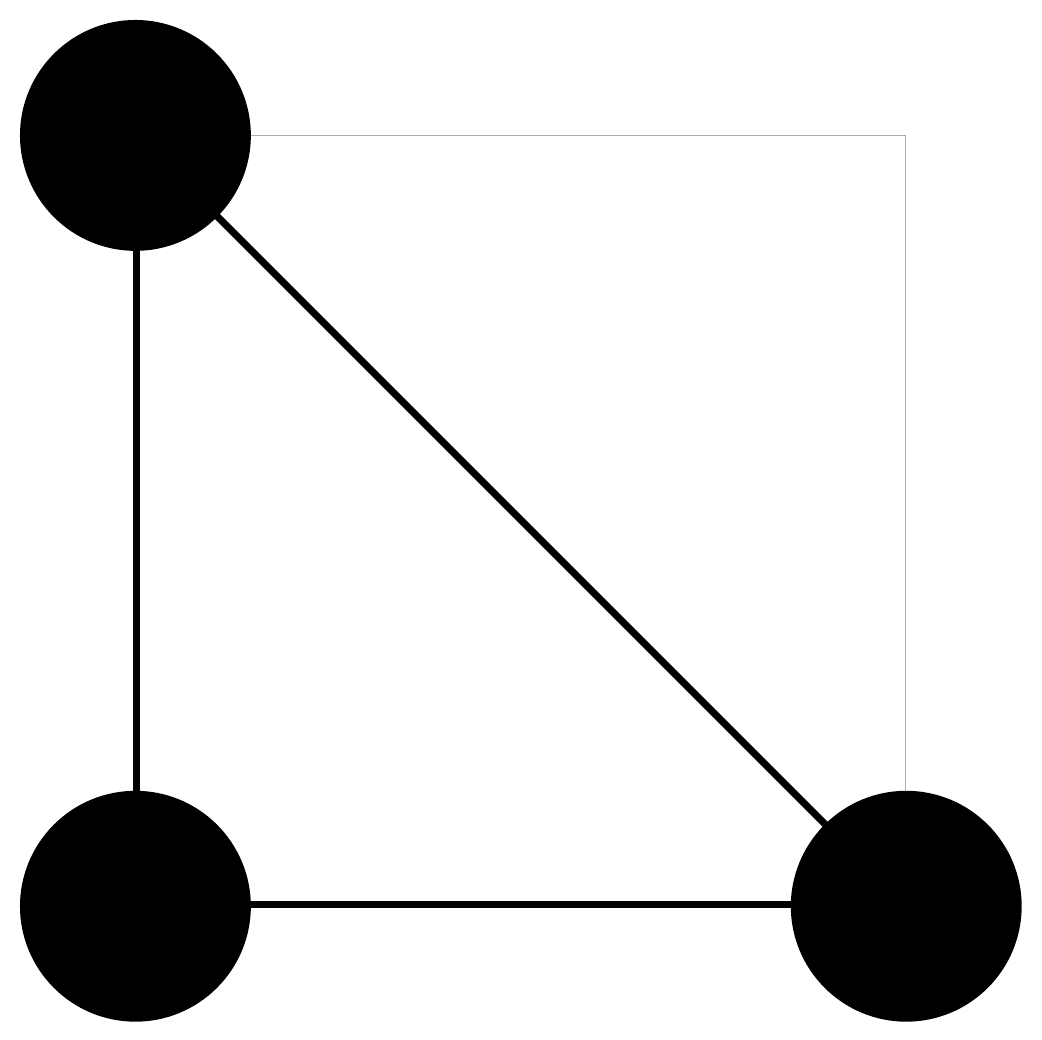} 
\times {1 \; 0 \choose 0 \; 2} = 
\includegraphics[height=1.4cm, trim=0cm 3cm 0cm 0cm ]{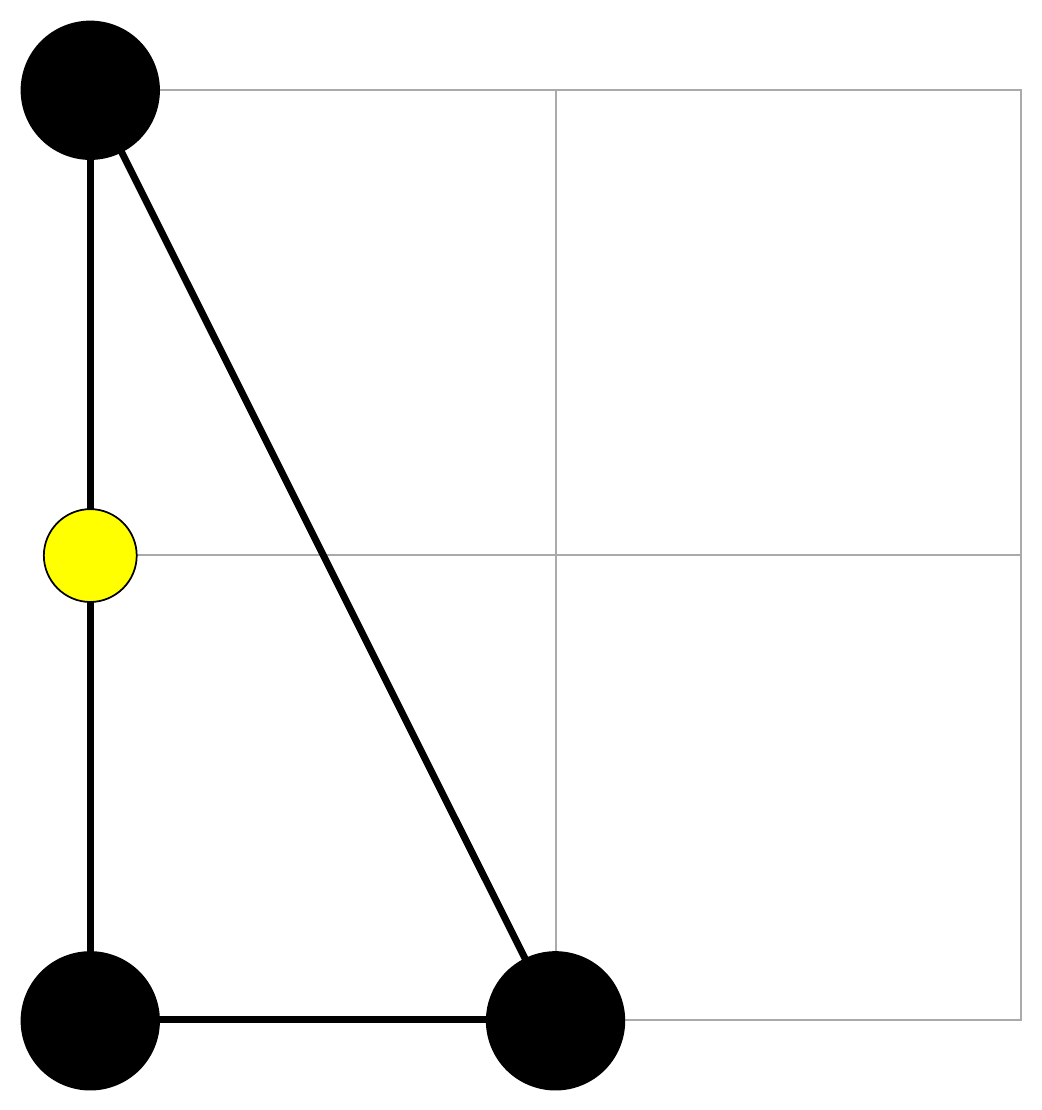} 
~~.
\label{eq:HNFZ2}
\eeq

The $2\times 2$ matrices one has to consider in order to cover all possible toric triangles of a given area are in Hermite Normal Form (HNF).

\subsection{Hermite Normal Form}
An upper triangular $2\times 2$ integer valued matrix of the form
\beq
M=
\left(
\begin{array}{cc}
a & b \\
0 & c
\end{array}
\right)~~,
\eeq
where $\det{M}=ac$ and $0\leq b < c$ is said to be in Hermite Normal Form (HNF). All $2 \times 2$ integer valued matrices can be written as the product of a HNF matrix and a matrix in $GL( 2 , \BZ)$. There are a finite number of integer valued matrices in HNF with any fixed determinant. Therefore, when generating triangles of a given area $|\Gamma|=\det{M}$, one only needs to consider this finite list of matrices in HNF in order to cover all distinct triangles.

\subsection{An Example - $\BC^3 / \BZ_3$}

Let us consider again the orbifolds of $\BC^3$ at order $|\Gamma|=3$. The HNF matrices of determinant $3$ and the corresponding toric triangles are
\beq
\underbrace{
\left( \begin{array}{cc}
1 & 0 \\
0 & 3
\end{array}\right)}_
{\includegraphics[width=1.7cm]{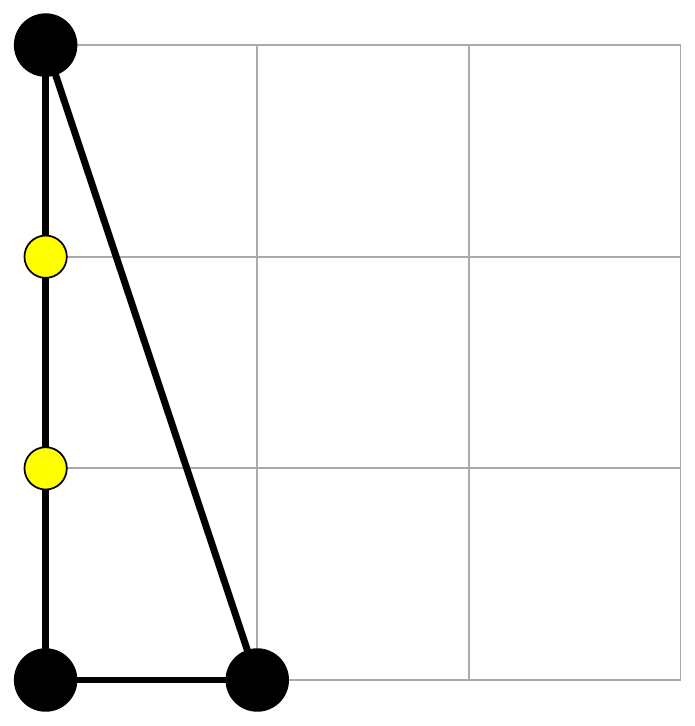}} ~~,~~
\underbrace{\left( \begin{array}{cc}
1 & 1 \\
0 & 3
\end{array}\right)}_
{\includegraphics[width=1.7cm]{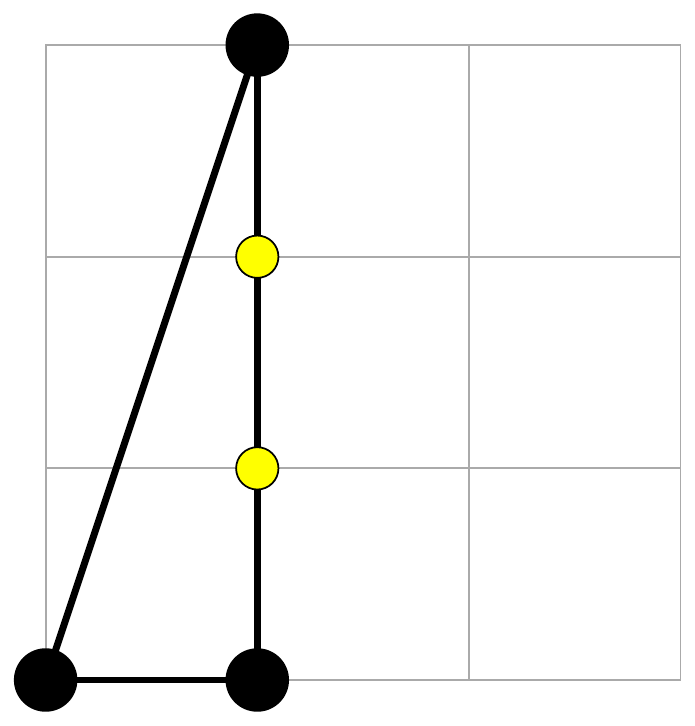}} ~~,~~
\underbrace{\left( \begin{array}{cc}
3 & 0 \\
0 & 1
\end{array}\right)}_
{\includegraphics[width=1.7cm]{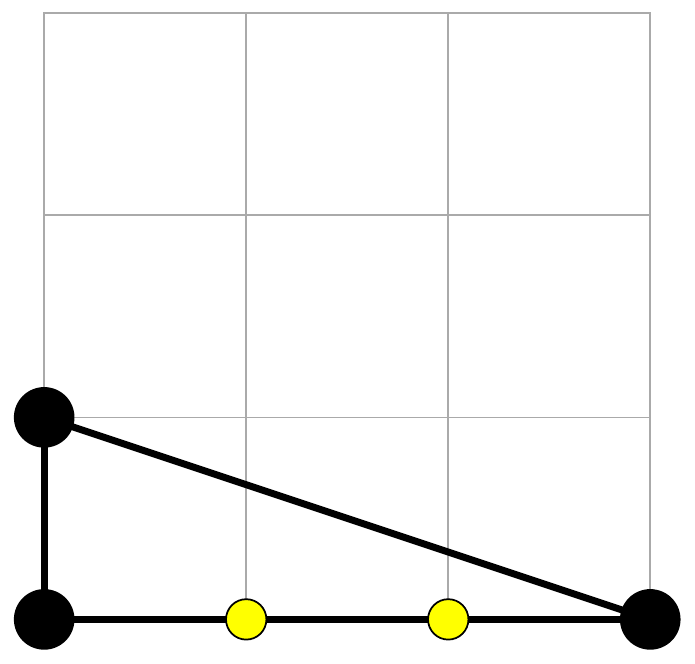}} ~~,~~
\underbrace{\left( \begin{array}{cc}
1 & 2 \\
0 & 3
\end{array}\right)}_
{\includegraphics[width=1.7cm]{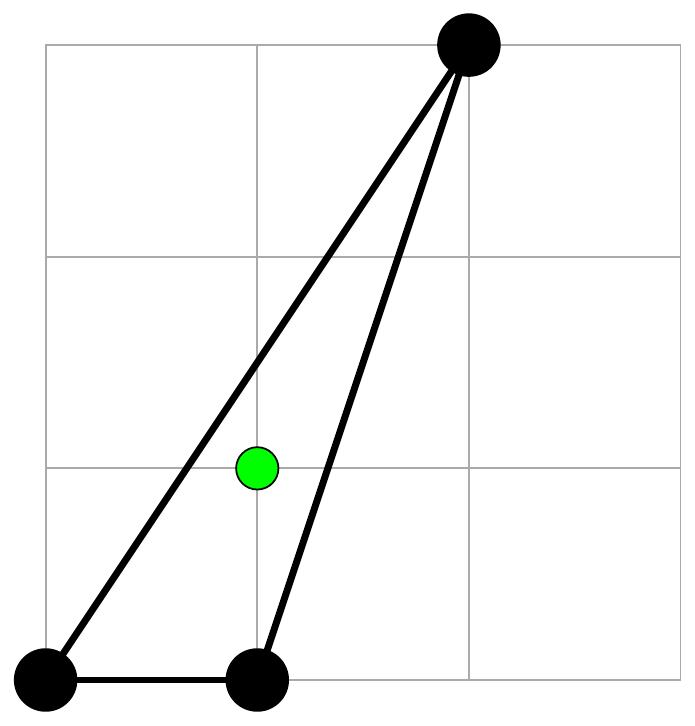}} ~~.
\label{HNFTriangles}
\eeq
Each of the triangles in \eqref{HNFTriangles} have an edge which is parallel to the x-axis because all $2 \times 2$ matrices in HNF have a lower left entry which is zero.
One observes that there are two distinct abelian orbifolds of $\mathbb{C}^3$ at order $|\Gamma|=3$, which are summarised in Table \ref{tab:3vecs}.

\section{Counting Orbifolds using Brane Tilings}
\label{s:TilingEnum}
A third way in which one can count abelian orbifolds of $\BC^3$ is by using an object called the brane tiling  \cite{Hanany:1997tb,Hanany:1998ru,HananyUranga98}. Brane tilings are periodic bipartite graphs on the plane. They are used to describe quiver gauge theories which are world-volume theories of a D3-brane probing a toric CY singularity. Table \ref{t0} shows the dictionary between a brane tiling and the quiver gauge theory.

\begin{table}[h!]
\begin{tabular}{|c|c|c|}
\hline
{\bf Brane Tiling} & {\bf String Theory} & {\bf Gauge Theory} \\ \hline 
Face & D5-branes & Gauge group \\ \hline
Edge between two & String stretched between D5-  & Bifundamental chiral multiplet\\
 faces             &branes branes through NS5 brane  &  \\ \hline
$k$-valent vertex & Region where $k$ strings & Interaction between $k$ chiral \\
 & interact locally.&multiplets, i.e. Superpotential \\
 && term of order $k$ \\ 
 \hline
\end{tabular}
\caption{Dictionary for translating between a brane tiling, string theory and gauge theory \cite{Hanany05}.}\label{t0}
\end{table}

Brane tilings formed from only hexagonal faces correspond to gauge theories whose moduli space is an abelian orbifold of $\BC^3$. The number of distinct faces or gauge groups in the corresponding quiver gauge theory is the order $|\Gamma|$ of the orbifold. Therefore, by counting all possible distinct hexagonal brane tilings formed by $|\Gamma|$ hexagons, one also counts abelian orbifolds of the form $\BC^3 / \Gamma$ \cite{Sloane97}.

\subsection{An Example - $\BC^3 / \BZ_3$}

Let us consider again the abelian orbifolds of $\BC^3$ at order $|\Gamma|=3$. Starting with $3$ distinct hexagons which we label from $1$ to $3$, one finds the following brane tiling constructions
\beq
\underbrace{
\includegraphics[height=3cm, trim=1cm 0cm 3cm 0cm]{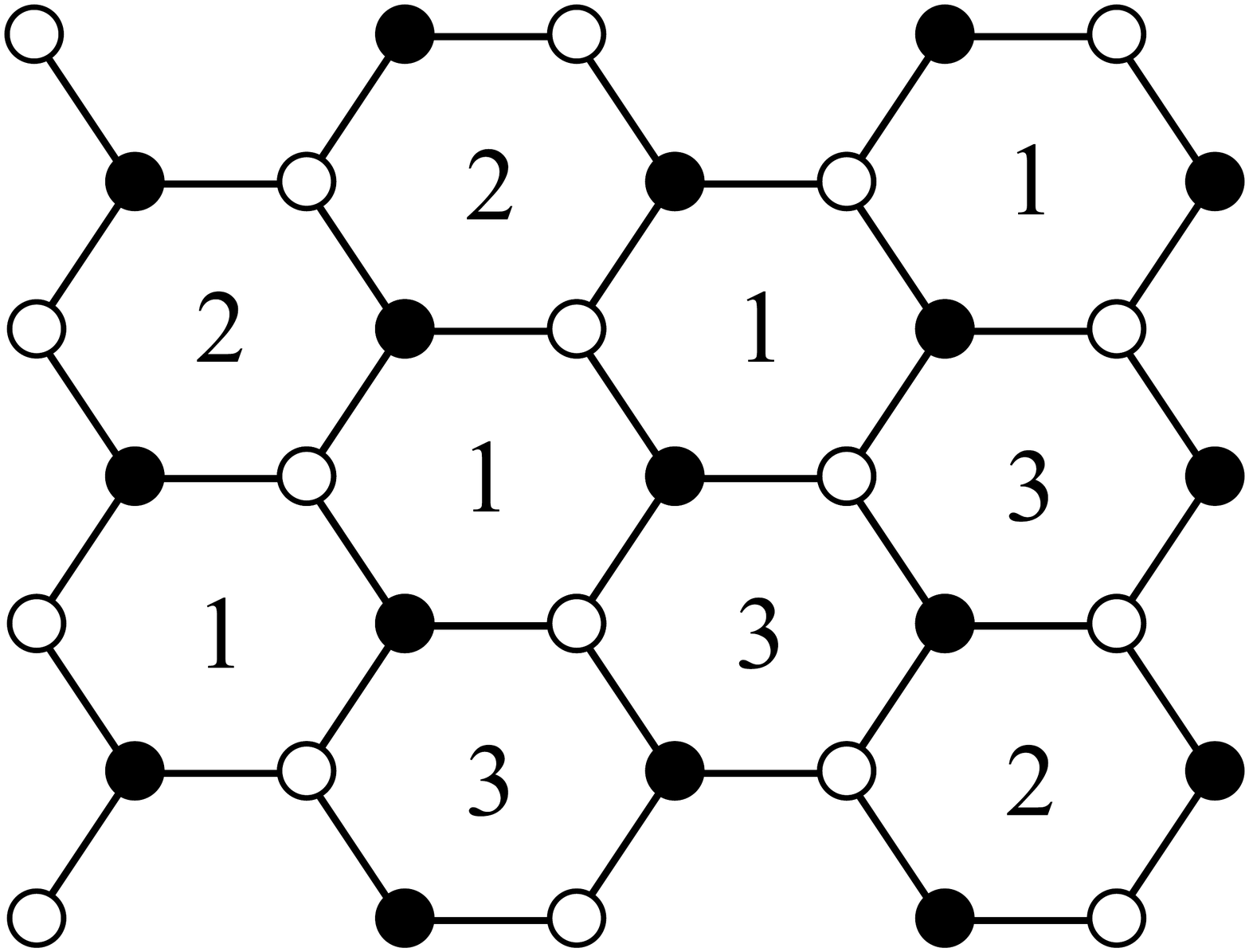}
\includegraphics[height=3cm, trim=1cm 0cm 3cm 0cm]{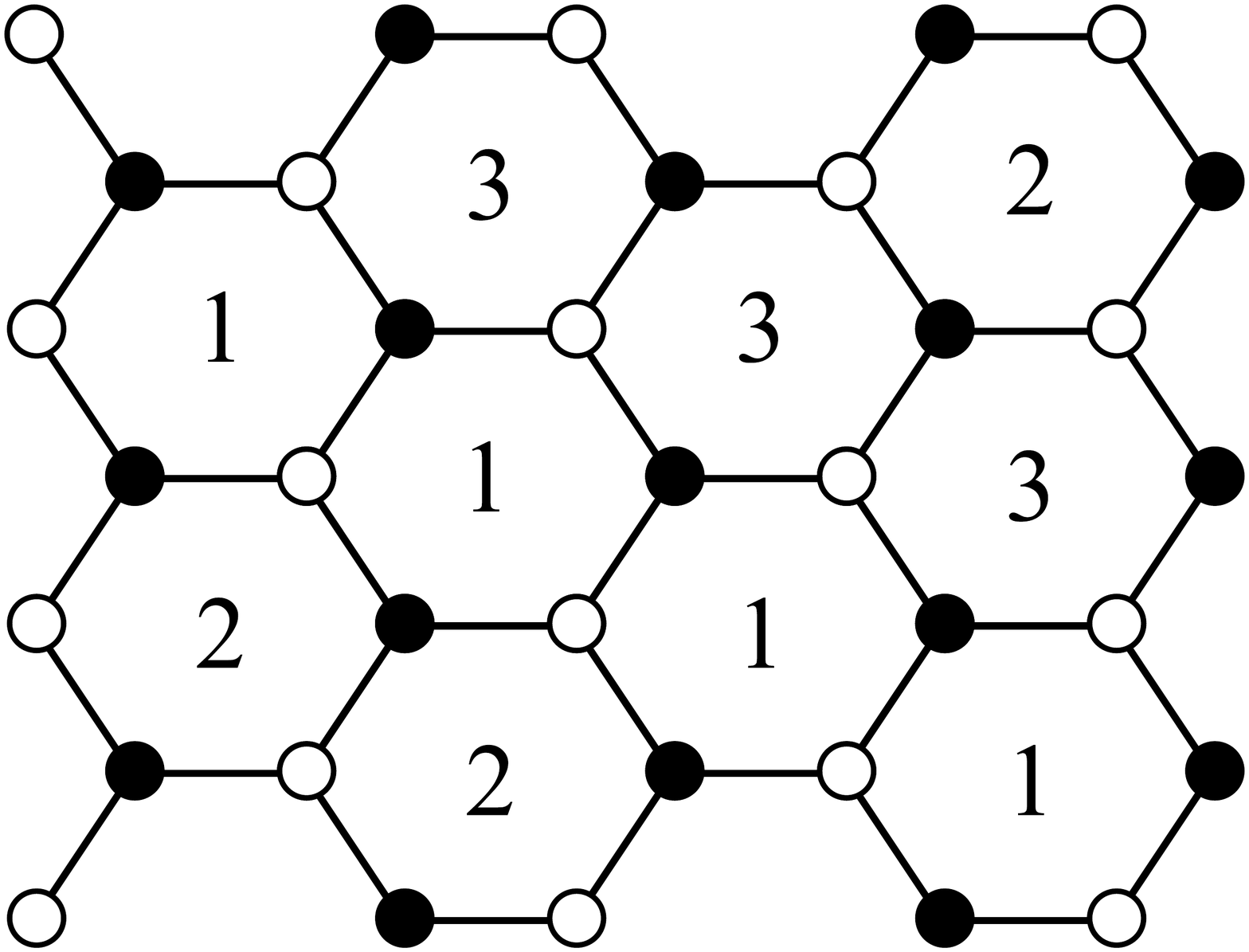}
\includegraphics[height=3cm, trim=1cm 0cm 3cm 0cm]{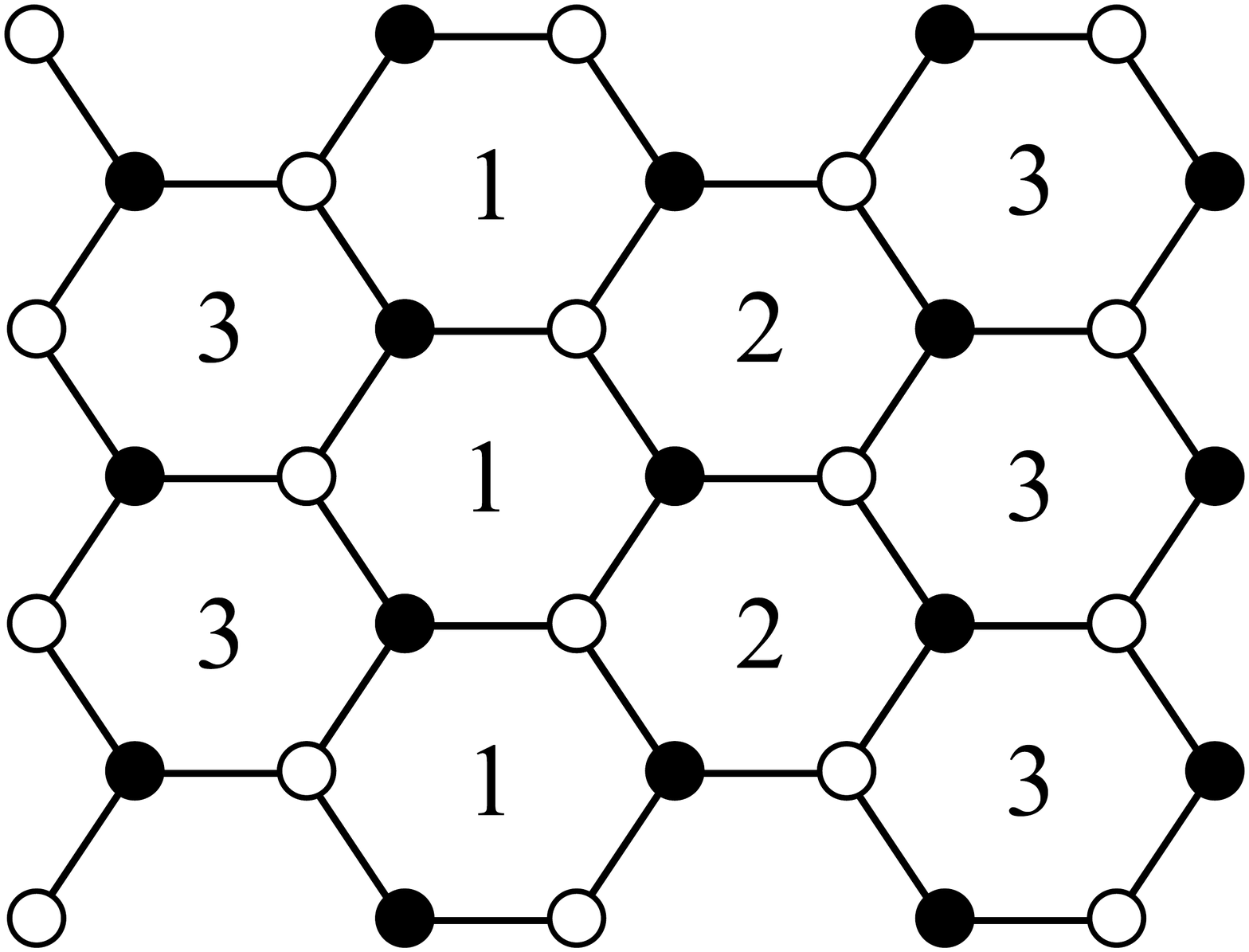}
}_{\mathbb{C}\times\mathbb{C}^2/\mathbb{Z}_3}
~~
\underbrace{
\includegraphics[height=3cm, trim=1cm 0cm 3cm 0cm]{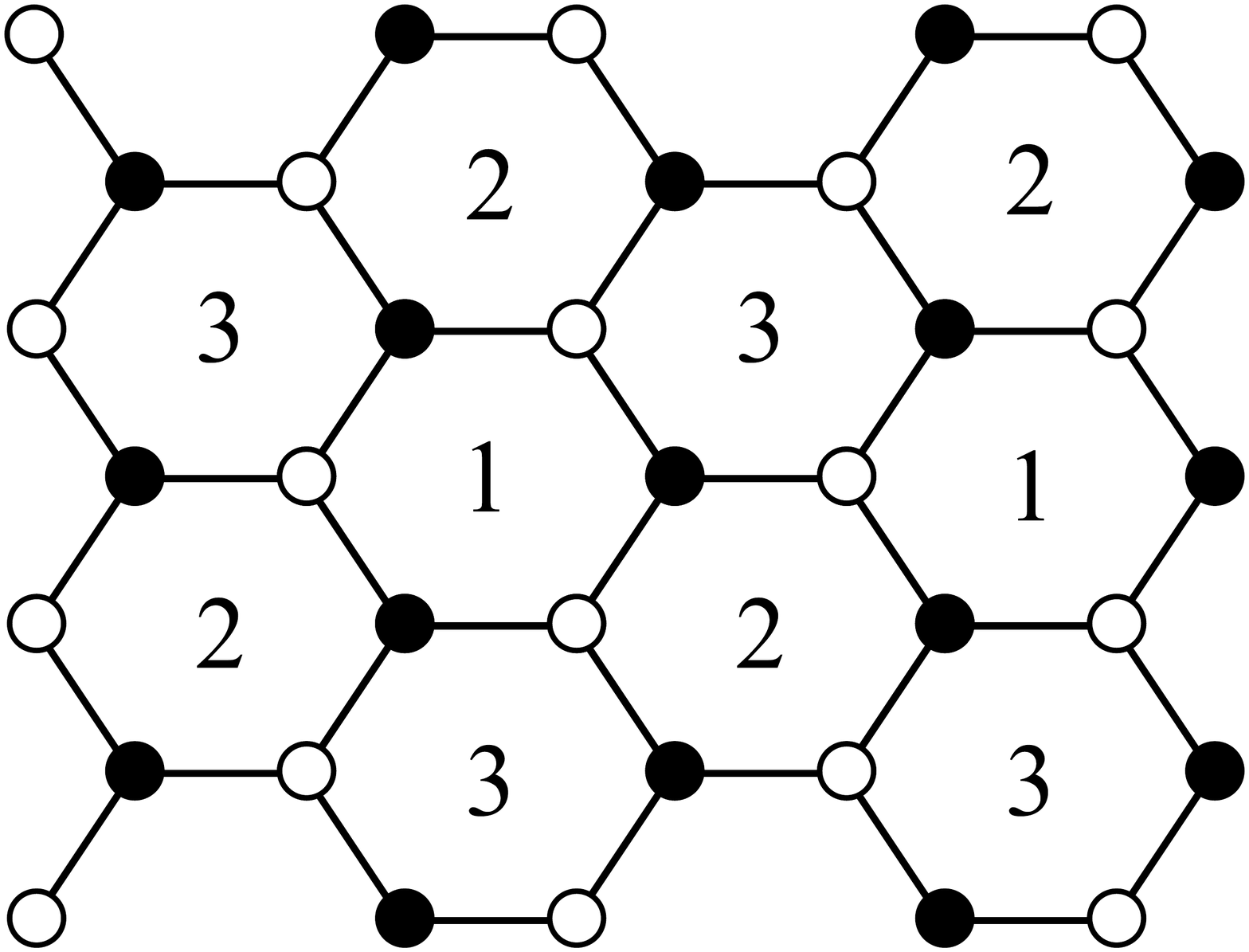}
}_{\mathbb{C}^3/\mathbb{Z}_3}
\eeq
where one observes that there are two distinct brane tilings which precisely corresponds to the orbifolds of the form $\mathbb{C}\times\mathbb{C}^2/\mathbb{Z}_3$ and $\mathbb{C}^3/\mathbb{Z}_3$ as summarized in Table \ref{tab:3vecs}.

%
%\subsection{Relationship Between Orbifold Action and Brane Tilings}
%In the 3 hexagon example, we have seen that there is a link between the hexagonal lattices and the tuple that we used to represent the orbifold action in Section \ref{s:OrbEnum}. This idea can be extended and used to count all abelian CY orbifolds of the form $\BC^3 / \Gamma$.\\
%For orbifolds that can be written as $\BC^3 / \BZ_n$, the 3-tuple that defines that orbifold corresponds to the way the label changes as we go from one face in the tiling to the next. This is shown clearly in \todo{figure}.\\
%For orbifolds that are true $\BC^3 / \BZ_n \times \BZ_m$ orbifolds (ie they can't be written as a $\BC^3 / \BZ_n$ orbifold), it is simplest to represent the labels on faces of the hexagon by two numbers. The way in which the first number changes from one face in the tiling to the next corresponds to the 3-tuple associated to the $\BZ_n$ action. The way in which the sceond number changes from face to face corresponds to the second 3-tuple (the one that defines the action of $\BZ_m$ on $\BC^3$. The way in which these actions work out is given in figure\\
%%%%%%%%%%%%%%%%%%%%%%%%%%%%%%%%%%%%%%%%%%%%%%%%%%%%%%%%%%%%%%%%%%%%%%%%%%%%%%%
\section{Explicit Counting}
\label{s:Counting}
The methods given above are used to count abelian orbifolds of $\BC^3$. These three methods are equivalent and give the same counting. Let the number of orbifolds of the form $\BC^3 / \Gamma$ at order $|\Gamma | = n$ be $f(n)$. The first 50 values of $f(n)$ are given in Table \ref{t:fn}.\\
\begin{table}[h!]
  \begin{tabular}{c|ccccccccccccccccccccccccc}
    $n$ &
 1 		&
 2			&
 3			&
 4			&
 5			&
 6			&
 7			&
 8			&
 9			&
 10		&
 11		&
 12		&
 13		&
 14		&
 15		&
 16		&
 17		&
 18		
 \\ \hline
    $f(n)$ &
 1			&
 1			&
 2			&
 3			&
 2			&
 3			&
 3			&
 5			&
 4			&
 4			&
 3			&
 8			&
 4			&
 5			&
 6			&
 9			&
 4			&
 8			 \end{tabular}   \begin{tabular}{c|ccccccccccccccccccccccc}
    $n$ &
 19		&
 20		&   
 21		&
 22		&
 23		&
 24		&
 25		&
 26		&
 27		&
 28		&
 29		&
 30		&
 31		&
 32		&
 33		&
 34		\\ \hline
    $f(n)$ 
    &
 5			&
 10	 &
 8			&
 7			&
 5			&
 15		&
 7			&
 8			&
 9			&
 13		&
 6			&
 14		&
 7			&
 15		&
 10		&
 10				
  \end{tabular}
  \begin{tabular}{c|ccccccccccccccccccccccc}
    $n$ &
 35		&
 36		&
 37		&
 38		& 
 39		&
 40		&
 41		&
 42		&
 43		&
 44		&
 45		&
 46		&
 47		&
 48		&
 49		&
 50				
	\\ \hline
    $f(n)$ 
    &
 10		&
 20		&
 8			&
 11 & 12		&
 20		&
 8			&
 18		&
 9			&
 17		&
 16		&
 13		&
 9			&
 28		&
 12		&
 17			
  \end{tabular}
\caption{The number of orbifolds of $\BC^3 / \Gamma $ for $n =1,\dots,50$}
\label{t:fn} 
\end{table}

By writing the sequence $f(n)$ in terms of a partition function $F(t) = \sum f(n) t^n$, one finds the formula \cite{HananyOrlando10}
\beq
F(t) = \sum_{m=1}^\infty \left[ \frac{1}{\left( 1 - t^m \right) \left( 1 + t^{2m} \right) \left( 1 - t^{3m} \right)} - 1 \right] 
~~.
\label{eq:GenFunction}
\eeq

%%%%%%%%%%%%%%%%%%%%%%%%%%%%%%%%%%%%%%%%%%%%%%%%%%%%%%%%%%%%%%%%%%%%%%%%%%%%%%%
\section{Conclusions and Extensions}

We reviewed three methods of counting abelian orbifolds of $\BC^3$. The methods involve the use of 3-tuples which encode the action of the quotienting group $\Gamma$, the use of the toric description of the abelian orbifolds, and the use of hexagonal brane tilings which encode the corresponding quiver gauge theory.

Two of these methods which use tuples and toric diagrams can be generalised to count any higher dimensional abelian orbifold of $\BC^d$ with $d > 3$ \cite{Orbs,RakNew}. One can extend the idea of a 3-tuple that defines an action of a cyclic group on $\BC^3$ to a $d$-tuple that defines the action of a cyclic group on $\BC^d$. It is also possible to use toric data to count orbifolds of $\BC^d$. For instance, to count the abelian orbifolds of $\BC^4$, one must count distinct tetrahedra in a $\BZ^3$ lattice of a volume $|\Gamma|$. Higher dimensional simplices must be considered to count orbifolds of $\BC^d$ for $d>4$. Currently, it is not well understood how to extend the idea of the brane tiling to describe and count all abelian orbifolds of $\BC^4$ in the context of Chern-Simons gauge theories. A promising solution may be provided by brane crystals \cite{M2Crystal} which may be able to describe all distinct abelian orbifolds of $\BC^4$.\\
%It is possible to consider orbifolds of other spaces, for instance orbifolds of the conifold. This is an interesting area and work has recently been done in this direction. The technology used to compute a closed form for the partition function that counts abelian orbifolds of $\BC^3$ may prove to be useful in the discussion of orbifolds of general complex space.\\
% Use this code if you wish to generate your bibliography with BibTeX;
% please replace first the string "demo" below with the name(s) of
% the BibTeX data base(s) you want to use.
% The resulting bibliography-output (the contents of the .bbl file)
% must be pasted into this file before submission.
% 
% \bibliographystyle{pss}
% \bibliography{demo}
% 
% Replace the following example bibliography with your references
% before submission:
\section*{Acknowledgements}
R.-K. S. likes to thank his parents for their encouragement and support. He would also like to thank David
Weir.

\end{document}